\documentclass[fleqn]{2017SCGE}
\newcommand{\angstrom}{{\rm \mathring A}}
\newcommand{\mgii}{\hbox{Mg\,{\sc ii}}}
\newcommand{\civ}{\hbox{C\,{\sc iv}}}
\newcommand{\citet}[1]{\cite{#1}}
\newcommand{\citep}[1]{\cite{#1}}
\def\citept[#1][#2]#3{[#1~\citen{#3}]}
\begin{document}

\ensubject{subject}

\ArticleType{Article}
\Year{2018}
\Month{??}
\Vol{??}
\No{??}
\DOI{10.1007/??}
\ArtNo{??}
\ReceiveDate{?? ??, 2018}
\AcceptDate{?? ??, 2018}

\title{New Clues to Jet Launching: The inner disks in radio loud quasars may be more stable}{New Clues to Jet Launching: The inner disks in radio loud quasars may be more stable}

\author[1,2]{Zhen-Yi Cai}{{zcai@ustc.edu.cn}}
\author[1,2]{Yu-Han Sun}{{sunyh92@mail.ustc.edu.cn}}
\author[1,2]{Jun-Xian Wang}{{jxw@ustc.edu.cn}}
\author[1,2,3]{\\Fei-Fan Zhu}{}
\author[4,6]{Wei-Min Gu}{}
\author[5,6]{Feng Yuan}{}

\AuthorMark{Cai Z Y}

\AuthorCitation{Cai Z Y, Sun Y H, Wang J X, et al}

\address[1]{CAS Key Laboratory for Research in Galaxies and Cosmology, Department of Astronomy, University of Science and Technology of China, Hefei 230026, China}
\address[2]{School of Astronomy and Space Science, University of Science and Technology of China, Hefei 230026, China}
\address[3]{Department of Astronomy and Astrophysics, 537 Davey Lab, The Pennsylvania State University, University Park, PA 16802, USA}
\address[4]{Department of Astronomy, Xiamen University, Xiamen, Fujian 361005, China}
\address[5]{Shanghai Astronomical Observatory, Chinese Academy of Sciences, 80 Nandan Road, Shanghai 200030, China}
\address[6]{SHAO-XMU Joint Center for Astrophysics, Xiamen University, Xiamen, Fujian 361005, China}

\contributions{Zhen-Yi Cai and Yu-Han Sun contributed equally to this work.}

\abstract{
Jet launching in radio loud (RL) quasars is one of the fundamental problems in astrophysics.
Exploring the differences in the inner accretion disk properties between RL and radio quiet (RQ) quasars might yield helpful clues to this puzzle.
We previously discovered that the shorter term UV/optical variations of quasars are bluer than the longer term ones, i.e., the so-called timescale-dependent color variation. This is consistent with the scheme that the faster variations come from the inner and hotter disk regions, thus providing a useful tool to map the accretion disk which is otherwise unresolvable. 
In this work we compare the UV/optical variations of RL quasars in SDSS Stripe 82 to those of several RQ samples, including those matched in redshift-luminosity-black hole mass and/or color-magnitude.
We find that while both RL and RQ populations appear bluer when they brighten, RL quasars potentially show a weaker/flatter dependence on timescale in their color variation.
We further find that while both RL and RQ populations on average show similar variation amplitudes at long timescales, fast variations of RL sources
appear weaker/smaller (at timescales of $\sim$ 25 -- 300 days in the observer's frame),
and the difference is more prominent in the $g$-band than in the $r$-band.
Inhomogeneous disk simulations can qualitatively reproduce these observed differences if the inner accretion disk of RL quasars fluctuates less based on simple toy models. Though the implications are likely model dependent, the discovery points to an interesting diagram that magnetic fields in RL quasars may be prospectively stronger and play a key role in both jet launching and the stabilization of the inner accretion disk.
}



\maketitle


\begin{multicols}{2}
	
\section{Introduction}

\subsection{Jet launching in quasars}

How relativistic jets are launched in various types of 
\Authorfootnote
\noindent astronomical objects is one of the fundamental questions in 
modern astrophysics, and remains unclear after decades of studies.
The current general consensus is that the jet energy is extracted from the central black hole (BH) spin \citept[e.g.,][]{BlandfordZnajek1977} and/or the surrounding accretion disk \citept[e.g.,][]{BlandfordPayne1982, Lynden-Bell2003}.

Quasars (with BH masses and accretion rates in the regime where a standard thin disk is expected)
probe different parameter space regions as compared with the low hard states of X-ray binaries and low luminosity radio galaxies. Therefore their jet launching mechanism might be different \citept[e.g.,][]{YuanNarayan2014}. 
Conventionally, the ratio of radio to optical flux density (i.e., the radio loudness $R$) is used to separate quasars into radio loud (RL) and radio quiet (RQ) populations. Though the fractions of RL quasars are usually discussed using different flux-limited samples and different optical/radio bands, they are generally around $\sim 10\%$ (e.g., 15\%-20\% estimated by Kellermann et al. \citep{Kellermann1989} and 8\%$\pm$1\% by Ivezi{\'c} et al. \citep{Ivezic2002}).
Why are there only a small fraction of RL quasars, and what is the key underlying difference between RL and RQ quasars?

Theoretically, the jet formation in RL quasars leaves imprints on their accretion disk emissions \citept[e.g.,][]{Zamaninasab2014}. Observationally, although the main characteristics
 of the global UV/optical spectral energy distributions (SEDs; i.e., the integrated disk emission) between these two populations are similar \citept[e.g.,][]{Elvis1994}, moderate SED/color differences between RL and RQ quasars are also reported \citept[e.g.,][]{Francis2000,Ivezic2002,Kotilainen2007,Labita2008,Shankar2016}.
Comparing the inner accretion disk emissions from RL and RQ quasars would provide useful clues on these differences, though it is always difficult as the disk is unresolvable.
Interestingly, it was found that RL quasars tend to have redder extreme-ultraviolet (EUV) radiation, compared with RQ ones \citept[e.g.,][]{Telfer2002a}.
Punsly \citep{Punsly2015} connected such UV deficit to magnetically arrested accretion in the innermost disk regions of RL quasars, which removes energy from the accretion flow as Poynting flux.
Meanwhile, such difference in the EUV slope can also be attributed to additional dust reddening \citep{Binette2005},  smaller Eddington ratios, and/or larger BH masses \citept[e.g.,][]{McLureJarvis2004} in RL quasars, as the observed spectral slope could be determined by many constituents.

In this work we present a useful probe to investigate how the inner accretion disks in RL quasars are distinct from those in RQ ones.

\subsection{Probing the accretion disk with color variation at various timescales}

Quasars are variable from radio to X-ray and gamma-ray. In UV/optical bands, where the disk emission dominates, the variation amplitude is
wavelength dependent, in a way that quasars generally show larger variation amplitudes in bluer bands \citept[e.g.,][]{Giveon1999,Hawkins2003,VadenBerk2004,MacLeod2010,Sakata2011,Schmidt2012,Ruan2014,Sun2014,Sun2015,Guo2016}. 
Since the variations across UV/optical continuum occur almost simultaneously in phase, with a time lag less than 1-2 days \citep{Giveon1999,Hawkins2003}, quasars normally appear bluer when they get brighter, which is the so-called ``bluer when brighter'' (BWB) trend \citept[e.g.,][]{Schmidt2012,Ruan2014,Sun2014,Guo2016}.

Recently, we discovered that the BWB trend is timescale-dependent for quasars in the Sloan Digital Sky Survey (SDSS) at all redshifts up to z $\sim$ 3.5 \citep{Sun2014}, i.e., the shorter term variations are even bluer than the longer term ones. Such timescale dependence has been confirmed down to the rest-frame EUV bands with the {\it GALaxy Evolution eXplorer} (GALEX) data \citep{Zhu2016}. This discovery can immediately rule out alternative explanations to the BWB trend, including host galaxy contamination and changes in global accretion rate, as these models intuitively imply timescale-independent color variations. More importantly, this discovery indicates that the shorter term variations come from the inner most regions of the accretion disk where the disk temperature is higher, while the longer term variations come from the outer disk regions with lower effective temperature. Using Monte-Carlo simulations, Cai et al. \citep{Cai2016} showed that a thermal-fluctuating accretion disk model (adjusted from Dexter \& Agol \citep{DexterAgol2011}) can well reproduce the observed timescale dependence in optical and UV bands. A highly interesting consequence from these studies is that one can use variations at different timescales to spatially ``resolve'' the disk emission.

We note that Ramolla et al. \citep{Ramolla2015} found that the $B/V$ color variation (in flux-flux space) of 3C 120, a broad line radio galaxy, shows rather weak timescale dependence, 
and our independent analyses of 3C 120 yield similar pattern in both flux-flux and mag-mag spaces \citep{Cai2016}. 
This rather weak timescale dependence of 
color variation in 3C 120 hints that RL sources may behave differently compared with SDSS Stripe 82 quasars which are mostly RQ, i.e., the accretion disks in RL sources have properties distinct from those of RQ ones. 

In this paper we utilize the light curves of SDSS Stripe 82 quasars to systemically explore whether RL quasars exhibit different timescale dependence in their color variation, compared with RQ ones.
We present the data and analyzing method in Section~\ref{sect:data_method}, and the main results in Section~\ref{sect:results}.  The implications of our results and discussions are given in Section~\ref{sect:discussion}.

\section{Data and Methodology}\label{sect:data_method}

\subsection{Data}\label{sect:data}

Following Sun et al. \citep{Sun2014}, we utilize the SDSS monitoring data of quasars in Stripe 82 presented by MacLeod et al. \citep{MacLeod2012}\footnote{\url{http://faculty.washington.edu/ivezic/macleod/qso_dr7/DB_QSO_S82.dat.gz}}, who provided re-calibrated $\sim$ 10 years long light curves in five SDSS bands ($ugriz$) for 9258 spectroscopically confirmed quasars in SDSS Data Release 7 (DR7). 
Each light curve generally contains $\sim$ 60 epochs.
As shown by Sun et al. \citep{Sun2014} and Zhu et al. \citep{Zhu2016}, large photometric uncertainties may produce significant bias to the observed color variation, 
especially at short timescales where the intrinsic variations are weak. 
In this work we focus on the $g$- and $r$-band light curves, which have the best photometric quality among the five bands (see Figure 2 of Sun et al. \citep{Sun2014}).
We exclude epochs with unphysical photometric magnitudes or with $g/r$-band photometric uncertainties greater than 0.1 mag, and only consider sources with more than 20 good epochs. 
Totally, 260 sources are rejected, including 21 sources with unphysical $g$- and $r$-band mean magnitudes in the catalog of MacLeod et al. \citet{MacLeod2012}.

To explore any difference of color variation between RL and RQ quasars, 
the resultant 8998 sources are further matched to the SDSS DR7 quasar catalog\footnote{\url{https://users.obs.carnegiescience.edu/yshen/BH_mass/data/catalogs/dr7_bh_Nov19_2013.fits.gz}} of Shen et al. \citet{Shen2011}
to get their redshift, $z$, BH mass, $M_{\rm BH}$, bolometric luminosity, $L_{\rm bol}$, the FIRST flag of radio FR type, and radio loudness, $R = f_{6 \rm cm}/f_{2500 \angstrom}$, where $f_{6 \rm cm}$ and $f_{2500 \angstrom}$ are the flux densities at rest-frame 6 cm and 2500~\AA, respectively.
There are three sources without counterpart within a matching radius of 18 arcsec, 88 sources without BH mass, three sources without bolometric luminosity, and 1321 sources without FIRST flag (out of FIRST footprints).
After rejecting those sources 
the final sample we derived contains 7185 RQ/undetected quasars ($R < 10$) and 416 RL sources ($R \geq 10$). 

\begin{figure}[H]
	\centering
	\includegraphics[width=0.4\textwidth]{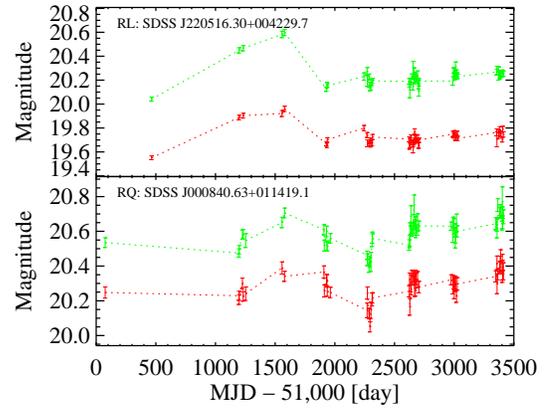}
	\caption{An illustration of the $g$- (green) and $r$-band (red) light curves of RL (top) and RQ (bottom) quasars randomly selected from our final sample.}
	\label{fig:example}
\end{figure}

Examples $g$- and $r$-band light curves of a RL and a RQ quasar, randomly selected from our final sample, are respectively illustrated in \cref{fig:example}. 
We note that those quality cuts on light curves do not produce any significant difference in the photometric sampling pattern between RL and  RQ quasars we would analyze in this work.

\subsection{Quantifying Color Variability at Various Timescales}\label{sect:cv}

For each quasar with simultaneous (or quasi-simultaneous) multi-band photometric observations (e.g., the SDSS $g$- and $r$-bands in this work) at two epochs, the color variation\footnote{According to this nomenclature, one may expect a definition like $\Delta {\cal C}(\tau) \equiv [m_g(t_2) - m_r(t_2)] - [m_g(t_1) - m_r(t_1)]  = \Delta m_g(\tau) - \Delta m_r(\tau)$ with $\tau = t_2 - t_1$. This is of course  similar to our definition, but $\Delta {\cal C}<(>)0$ only indicates an object becomes bluer (redder) without any further information on the change of magnitude. Instead, by adopting the ratio of magnitude difference to define the color variation as what we have done, one can easily get an object becomes bluer when it brightens if $0 < \Delta C_{rg} < 1$. For example, a 1.0 mag brightening in the $g$-band and a 0.9 mag brightening in the $r$-band (i.e., an object getting bluer when brightening) would make both the numerator and denominator of eq.~(\ref{eq:cv}) negative, but the denominator more negative, so one would end up with a positive number less than 1. At long timescales the same object may change instead by 1.0 and 0.95 mag in the $g$- and $r$-bands, so it is not as ``blue'' and the ratio increases.}
between each epoch pair can be defined as 
\begin{equation}\label{eq:cv}
\Delta C_{rg}(\tau) \equiv \frac{\Delta m_r (\tau)}{\Delta m_g (\tau)} = \frac{m_r(t_2)-m_r(t_1)}{m_g(t_2)-m_g(t_1)},
\end{equation} 
where $m_j(t_k)$ is the observed magnitude in $j$-band at epoch $t_k$, $\Delta m_j(\tau)$ is the difference of $j$-band magnitude between two epochs with time interval of $\tau$, and $\tau = t_2 - t_1$ refers to the timescale of the variation ($t_2 > t_1$).
Then, for a large sample of quasars, the $\Delta C_{rg}$ averaged in each timescale bin describes the ensemble color variation as a function of timescale, $\tau$. 

\noindent Unambiguously, $\Delta C_{rg} > 0$ defines quasars become brighter or dimmer simultaneously in both bands, and then $\Delta C_{rg} <$ ($>$) 1 indicates that quasars get bluer (redder) when they brighten. 
In this work, we nominate the short/long term as the smaller/larger time interval, $\tau$. Therefore, the short term variation is bluer than the long term one would mean that $\Delta C_{rg}$ is smaller with decreasing time interval, $\tau$.

Firstly, we exclude epoch pairs for which the time lags are less than 25 days, below which the intrinsic quasar variations are too weak compared with the photometric noise.
We also set a 3$\sigma$ cut that rejects epoch pairs, in which the magnitude variations in the $g - r$ space are statistically insignificant, to further reduce the effect of photometric uncertainties on the measurement of color variation \citep{Sun2014}.

Secondly, we consider only epoch pairs with $\Delta C_{rg} > 0$, and using its ``median'' value to describe the ensemble color variation, rather than the arithmetic/geometric mean one in order to avoid being biased by those extreme values. 
Following is how the ``median'' of $\Delta C_{rg}$ is defined after deepening the understanding of the statistics of $\Delta C_{rg}$.

\begin{figure}[H]
	\centering
	\includegraphics[width=0.4\textwidth]{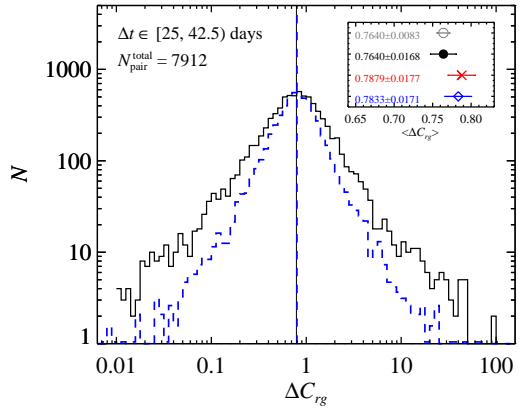}
	\caption{The histogram distributions of the observed $\Delta C_{rg}$ for the real RL quasar sample (black solid line) and the simulated $\Delta C_{rg}$ (blue dashed line; normalized at the peak of the solid histogram; see text for the detailed simulation in order to consider the effect of photometric uncertainty) in the timescale bin of 25 -- 42.5 days. 
	Note a few outliers with extreme values. 
	The vertical black solid and blue dashed lines show the standard medians of the real and simulated distributions, respectively. 
	The insert draws several ``averages'' and their uncertainties of the real $\Delta C_{rg}$ derived through different approaches. 
	From top to bottom, they are the geometric mean with $\sigma$/$\sqrt{N-1}$ uncertainty (grey open circle; but note this uncertainty is obviously underestimated),
	the geometric mean with bootstrap uncertainty (black filled circle), the standard median with bootstrap uncertainty (red cross), and the revised median with bootstrap uncertainty (blue open diamond; see Section~\ref{sect:cv} for details). 
	}
	\label{fig:hist}
\end{figure}

As an example, the black solid histogram in \cref{fig:hist} shows the real distribution of $\Delta C_{rg}$ in the shortest timescale bin of 25 -- 42.5 days for our RL quasar sample, providing a total of 7912 epoch pairs with $\Delta C_{rg} > 0$. 
A clear scatter of $\Delta C_{rg}$ is seen, and note a few epoch pairs have very small or large  $\Delta C_{rg}$.

The observed scatter of $\Delta C_{rg}$, however, can only be partially attributed to the photometric uncertainty, indicating an 

\noindent intrinsic scatter of $\Delta C_{rg}$.
Here, the scatter due to the photometric uncertainty is retrieved using simulations: for all epoch pairs $({\Delta m_g}, {\Delta m_r})$ in the shortest timescale bin, starting from the observed ${\Delta m_g}$, we calculate  the expected ${\Delta m_r}$ assuming a fixed 
value of $\Delta C_{rg}$ (i.e., the vertical black solid line in \cref{fig:hist} for the median value of the black solid histogram). Gaussian photometric uncertainties are then added to ${\Delta m_g}$ and ${\Delta m_r}$. The distribution of the simulated $\Delta C_{rg}$ is over-plotted in 
\cref{fig:hist} as the blue dashed histogram and its median as the vertical blue dashed line. We note here that for this shortest timescale bin, where the intrinsic flux variation is weakest among all bins, though the photometric uncertainty produces a clear scatter of $\Delta C_{rg}$, it has negligible effect to the median value. 

The difference between the real and simulated distributions of ${\Delta C_{rg}}$ implies that there is an extra intrinsic scatter of $\Delta C_{rg}$, which
may be attributed to the broad distribution of the quasar properties in our sample, including BH mass, luminosity, redshift, contaminations from the host galaxy and/or emission lines to the broadband photometries, and so on. Studying the exact nature of such intrinsic scatter is beyond the scope of this work. 

In this study, we would focus on the average of $\Delta C_{rg}$, which provides an ensemble description to the color variation of quasar samples. 
As the arithmetic mean is obviously an improper choice considering the quasi-log-normal distribution of $\Delta C_{rg}$, 
the geometric mean is plotted in the insert of \cref{fig:hist} as the gray open (or black filled) circle,
while the red cross draws the traditional median of $\Delta C_{rg}$.

There are essential facts we need to consider before calculating the statistical uncertainty of the average. 
Firstly, photometric measurement at one epoch may contribute to different epoch pairs, making these epoch pairs statistically dependent.
Secondly, since quasars show an intrinsic scatter in their color variation, the multiple epoch pairs from a single quasar, even if they are based on completely independent photometric measurements, can not be considered as statistically independent to each other.
For such datasets, the standard estimation of uncertainty, $\sigma/\sqrt{N^{\rm total}_{\rm pair}-1}$, where $\sigma$ and $N^{\rm total}_{\rm pair}$ are the standard deviation and total number of $\Delta C_{rg}$, would significantly underestimate the confidence range of the geometric mean (cf. the grey open circle in the insert of \cref{fig:hist}, where $N^{\rm total}_{\rm pair} = 7912$ for the aforementioned timescale bin). 

Fortunately, epoch pairs from different quasars are independent to each other, and the uncertainty of the average can be properly obtained through bootstrapping the quasar sample, i.e., measuring the standard deviation of simulated averages from 1000 randomly bootstrapped samples (cf. the error bars of the black filled circle, the red cross, and the blue diamond in the insert of \cref{fig:hist}).
While the averages from different approaches are similar (see the insert of \cref{fig:hist}), 
we confirm that the traditional median approach generally yields slightly smaller uncertainty, compared with the geometric mean.
This is because the median is insensitive to outliers, thus the traditional median is a better choice here compared with the geometric mean. 

\begin{figure}[H]
	\centering
	\includegraphics[width=0.4\textwidth]{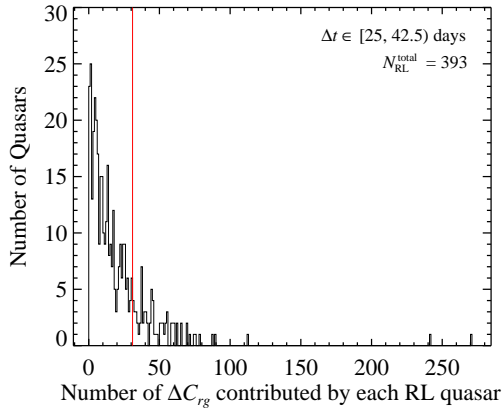}
	\caption{The histogram distribution of the number of $\Delta C_{rg}$ contributed by each RL quasar in the timescale bin of 25 -- 42.5 days. The vertical line plots the 80th percentile of source number, $N_{80\%}$, to the right of which 
	$\simeq 20\%$ quasars contribute to $\simeq 54\%$ of the total number of $\Delta C_{rg}$.
	}
	\label{fig:Number}
\end{figure}

As quasars have different variation amplitudes, each quasar does not contribute equally to a given timescale bin since we have filtered out those epoch pairs with insignificant variations. 
In \cref{fig:Number}, we plot the histogram distribution of the number of $\Delta C_{rg}$ contributed by each RL quasar in the shortest timescale bin of 25 -- 42.5 days. 
There are a total of 7912 positive $\Delta C_{rg}$, contributed by a total of 393 RL quasars in this timescale bin.
We note that a few quasars provide much more $\Delta C_{rg}$,
compared with the bulk of the sample. 
For instance, in \cref{fig:Number}, the vertical line indicates the 80th percentile of source number, $N_{80\%} = 31$, for this timescale bin.
To the right of the vertical line, 
79 ($\simeq 20\%$) quasars contribute 4251 ($\simeq 54\%$) $\Delta C_{rg}$ in this timescale bin.
Therefore, the standard median of $\Delta C_{rg}$ could thus be biased toward those a few quasars. 

To reduce such bias, we assign a lower weight of $N_{80\%}/N_i$ to each $\Delta C_{rg}$ from those top 20\% quasars with $N_i \ge N_{80\%}$ in each timescale bin, 
where $N_i$ is the number of $\Delta C_{rg}$ the $i$th quasar contributes.
A weight of unity is given to all $\Delta C_{rg}$ 

\noindent from the rest 80\% quasars. 
A weighted median of $\Delta C_{rg}$ can then derived.
As expected, in \cref{fig:hist}, we see that this revised median (blue open diamond) is slightly smaller, as well as its uncertainty (by a factor of $\simeq 3\%$), compared with the traditional median (red cross). 

We conclude that the statistic of the revised median of $\Delta C_{rg}$ is the best among those approaches we have discussed above. 
Hereafter, we present results based on this revised median (or the nominated average of $\Delta C_{rg}$, i.e., $\langle \Delta C_{rg} \rangle$, means this revised median). 
We stress that using the traditional median yields consistent results, and does not alter any of the scientific conclusions presented in this work.

\section{Results}\label{sect:results}

\begin{figure}[H]
	\centering
	\includegraphics[width=0.4\textwidth]{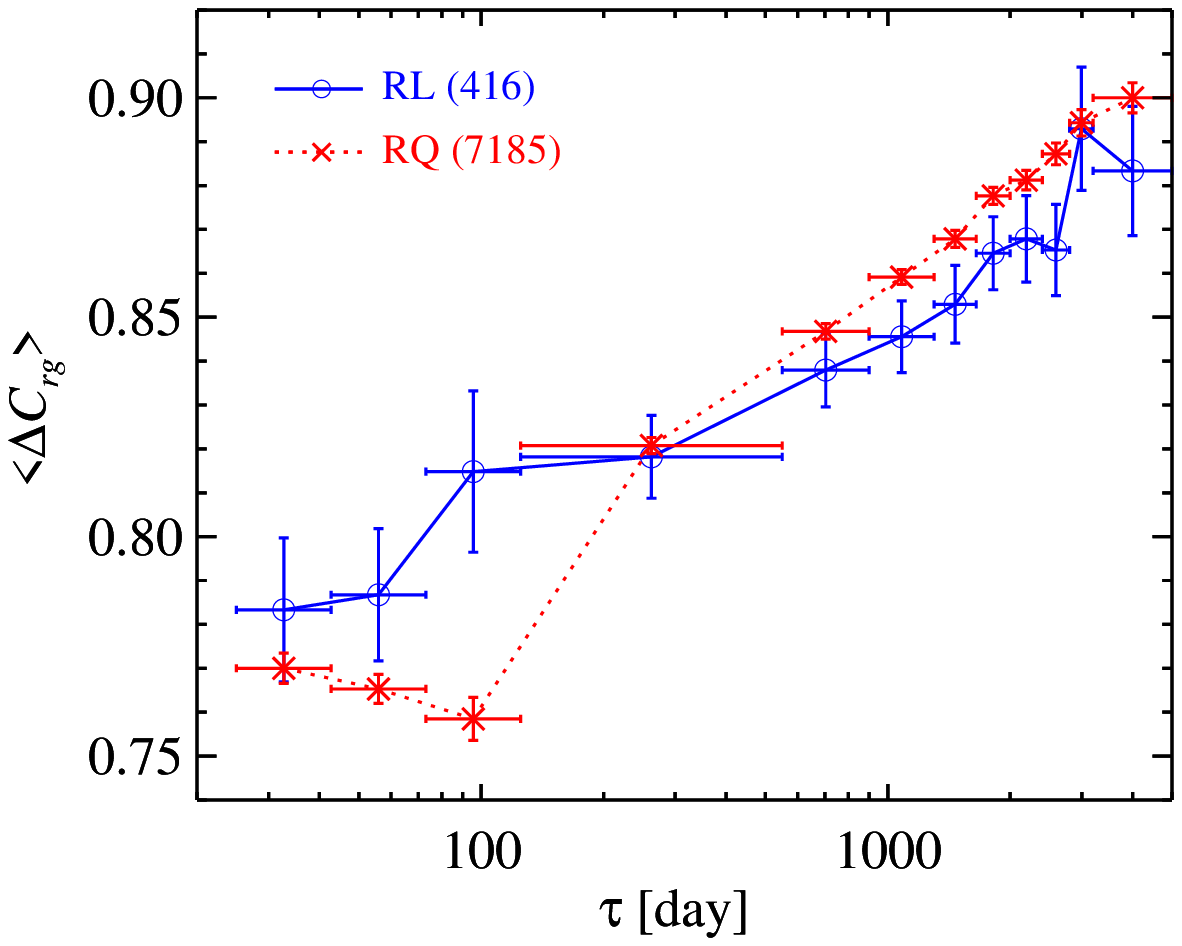}
	\includegraphics[width=0.4\textwidth]{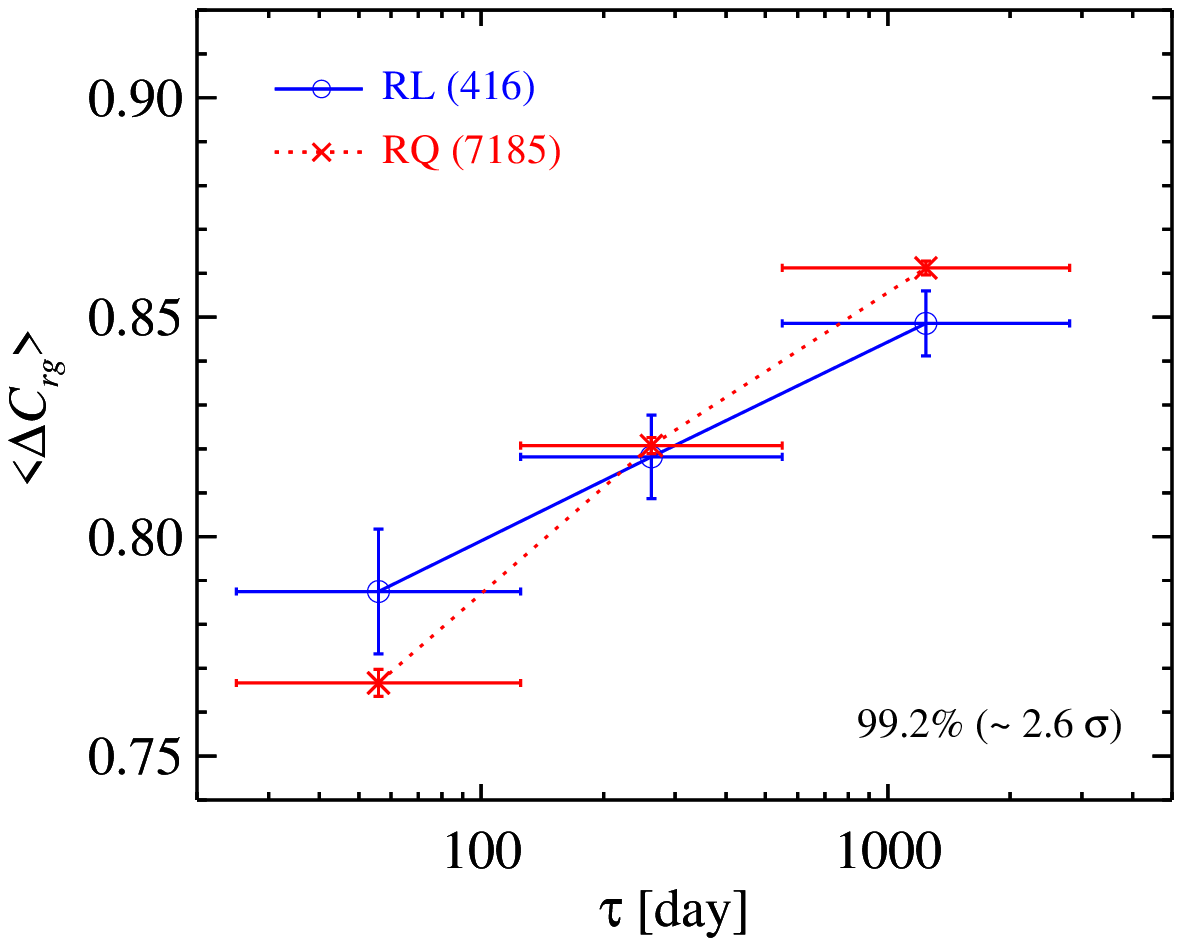}
	\caption{The top panel shows the $r$- versus $g$-band color variations, $\langle \Delta C_{rg} \rangle$, for the whole RL (blue open circles) and RQ (red crosses) samples as a function of timescale, $\tau$, in the observer's frame. The color variation of RL quasars shows a weaker/flatter timescale dependency with a confidence level of $\simeq 99.2\%~(\sim 2.6\sigma)$ estimated through bootstrapping. The total source number of each sample is nominated within the corresponding bracket at the top-left corner of each panel. (see Section~\ref{sect:results})
	}
	\label{fig:cv_whole_sample}
\end{figure}

In the top panel of \cref{fig:cv_whole_sample}, we plot the ensemble $\langle \Delta C_{rg} \rangle$ for our whole RL and RQ quasar samples as a function of timescale, $\tau$, in the observer's frame\footnote
{Note analyzing in the observer's frame ensures that data in each timescale bin equally come from the same sources for our sample.
Contrarily using rest-frame timescale bin would be biased: with the data point at longest rest timescale bin dominated by low-$z$ sources, and vice versa. Note that there are gaps in the observed timescale coverage, the bias would be even more complicated.}.
Here $\Delta C_{rg} < $ ($>$) 1 indicates that quasars get bluer (redder) when they brighten.

Similar to Sun et al. \citet{Sun2014}, both RL and RQ samples show the BWB trend at all timescales,
in a way that the shorter term variations are even bluer than the longer term ones.
However, it is interesting that, compared with RQ ones, the color variation of RL sample shows a weaker/flatter timescale dependency, which will be further investigated in the following.

The errorbars in the top panel of \cref{fig:cv_whole_sample} are obtained through bootstrapping the corresponding samples.
Note that the errorbars of the data points at various timescales in \cref{fig:cv_whole_sample}
are not statistically independent to each other. Instead, they are coupled since a given observation at a certain epoch 

\noindent will contribute to various timescale bins.
To assess the statistical significance of the difference in the $\langle \Delta C_{rg} \rangle$ -- $\tau$ relation between the RL and RQ quasars, we again adopt the bootstrap method to compare the ratio of the $\langle \Delta C_{rg} \rangle$ at shorter timescales to that at longer timescales. 
For each bootstrapped RL or RQ sample, we calculate $\langle \Delta C_{rg} \rangle_{25-125\rm days}$ over the three shortest timescale bins of 25-125 days, and $\langle \Delta C_{rg} \rangle_{550-2800\rm days}$ over the six longer timescale bins of 550-2800 days. 
The longest two timescale bins are excluded as there are much fewer epoch pairs in these bins. 
Taking 10000 times of bootstrapping, we generate 10000 pairs of $(RC^{\rm RL}, RC^{\rm RQ})$ for the RL and RQ samples, respectively, where the significance or the steepness of color variation with increasing timescale, $RC \equiv \langle \Delta C_{rg} \rangle_{550-2800\rm days}/\langle \Delta C_{rg} \rangle_{25-125\rm days}$, is defined for both samples nominated by the corresponding superscripts.
As shown in the bottom panel of \cref{fig:cv_whole_sample},  99.2\% of the 10000 

\begin{figure}[H]
	\centering
	\includegraphics[width=0.4\textwidth]{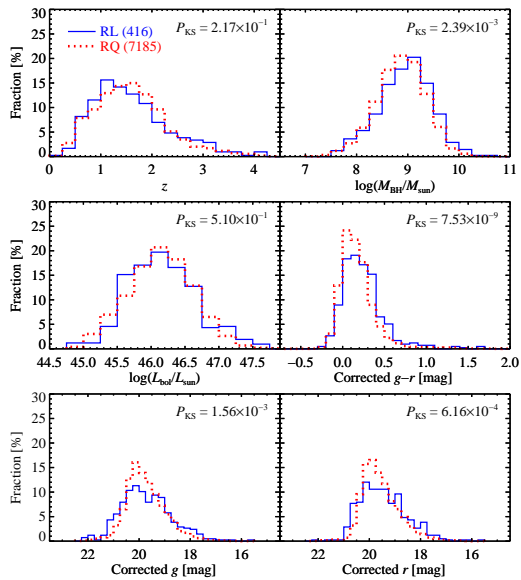}
	\caption{The distributions of redshift, BH mass, bolometric luminosity, Galactic extinction-corrected $g-r$ color, $g$-band apparent magnitude, and $r$-band apparent magnitude of the whole RL (blue solid line) and RQ (red dotted line) samples. According to the K-S probability, $P_{\rm KS}$, the RL quasars have comparable properties of redshift and luminosity, but slightly larger BH mass, redder color, and distinct apparent magnitudes, to those of RQ ones. 
	The total source number of each sample is nominated within the corresponding bracket at the top-left corner of the first panel.
	}
	\label{fig:dis_whole_sample}
\end{figure}

\noindent simulations yield smaller $RC$ from the RL quasars compared with the RQ ones, 
indicating that the RL quasars may show a weaker/flatter timescale dependency.

MacLeod et al. \citet{MacLeod2010,MacLeod2012} complied these spectroscopically confirmed quasars in Stripe 82, most ($\sim 8974$) of which were in the SDSS DR5 Quasar Catalog \citep{Schneider2007} and the remaining were confirmed as DR7 quasars \citep{Abazajian2009}. Following Richards et al. \citet{Richards2002a}, the bulk of these flux-limited quasars (with different flux limits for different redshift ranges) were selected based on their location in multi-dimensional SDSS color and magnitude spaces, complemented with the FIRST radio sources \citep{Becker1995}.
Due to the complexity of quasar selection in the Stripe 82, our whole RL and RQ samples may have different distributions in redshift, BH mass, Eddington ratio, color, and apparent magnitude,
which
may bias the aforementioned  difference between the two populations. In the following we explore these possible biases by constructing several sub-samples of RL and RQ sources matched in redshift-BH mass-bolometric luminosity (see Section~\ref{sect:resampled_z_m_Lb}) or/and color-magnitude (see Section~\ref{sect:resampled_z_gr_mg}).

\subsection{Sub-samples matched in $z$-$M_{\rm BH}$-$L_{\rm bol}$}\label{sect:resampled_z_m_Lb}

\cref{fig:dis_whole_sample} plots the distributions of redshift, BH mass, bolometric luminosity, $g-r$ color, and $g/r$ apparent magnitudes for the whole RL and RQ samples. All magnitudes and colors are corrected for the Galactic extinction.
Based on the Kolmogorov-Smirnov test (hereafter, shortly the K-S test),
we see no statistical difference between the two populations in redshift and bolometric luminosity, while RL quasars show slightly larger BH mass and redder $g-r$ color, consistent with previous studies \citept[e.g.,][]{Ivezic2002,MetcalfMagliocchetti2006,Shen2009,Kimball2011}.

In order to test whether the differences in the distributions of BH mass and/or color between RL and RQ quasars lead to the divergence of color variation shown in \cref{fig:cv_whole_sample},
we need to build samples with such properties matched. 

Considering the bins of $z$, $M_{\rm BH}$, and $L_{\rm bol}$ plotted in \cref{fig:dis_whole_sample}, 
we first construct sub-samples for RL and RQ sources by including only sources in those three-dimensional bins which contain both RL and RQ sources. 
The yielded RL and RQ sub-samples contain 367 and 4582 distinct sources.
We further adopt an acceptance-rejection sampling with replacement, to resample the RQ sub-sample to ensure it has the same three-dimensional distributions of redshift, BH mass, and bolometric luminosity as those of the RL sub-sample.
After resampling, the RQ sub-sample would contain a few duplicate sources, and each resampling for the RQ sub-sample will be slightly different, but the resultant color variations are similar thanks to the fact that the parent RQ sample contains numerous sources. Therefore, we only consider a single resampling of the RQ sub-sample, as shown in \cref{fig:dis_resampled_z_m_Lb}.
After being matched in $z$-$M_{\rm BH}$-$L_{\rm bol}$, the distributions of apparent magnitudes for both sub-samples are similar, while the difference between their $g-r$ color distributions is weaker but still significant. The latter will be further discussed in the next subsection.

\cref{fig:cv_resampled_z_m_Lb} compares the timescale-dependent color variations between the RL and resampled RQ sub-samples matched in $z$-$M_{\rm BH}$-$L_{\rm bol}$, and the difference between the two populations remains at a confidence level of 98.8\%.

\subsection{The effect of different $g-r$ color}\label{sect:resampled_z_gr_mg}

\begin{figure}[H]
	\centering
	\includegraphics[width=0.39\textwidth]{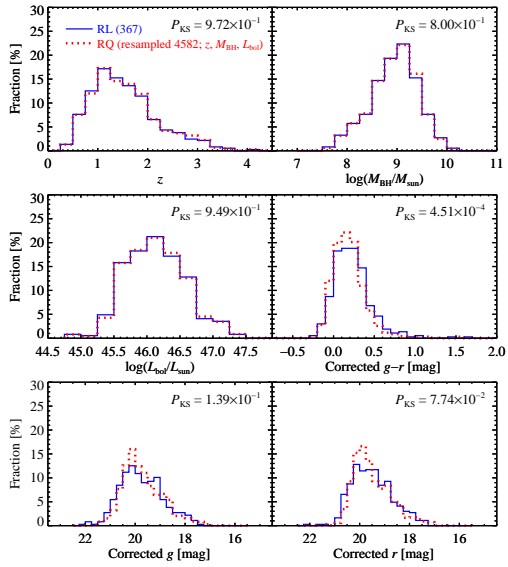}
	\caption{Same as \cref{fig:dis_whole_sample}, but for the RL and resampled RQ sub-samples matched in $z$-$M_{\rm BH}$-$L_{\rm bol}$. 
	(see Section~\ref{sect:resampled_z_m_Lb})
	}
	\label{fig:dis_resampled_z_m_Lb}
\end{figure}

\begin{figure}[H]
	\centering
	\includegraphics[width=0.39\textwidth]{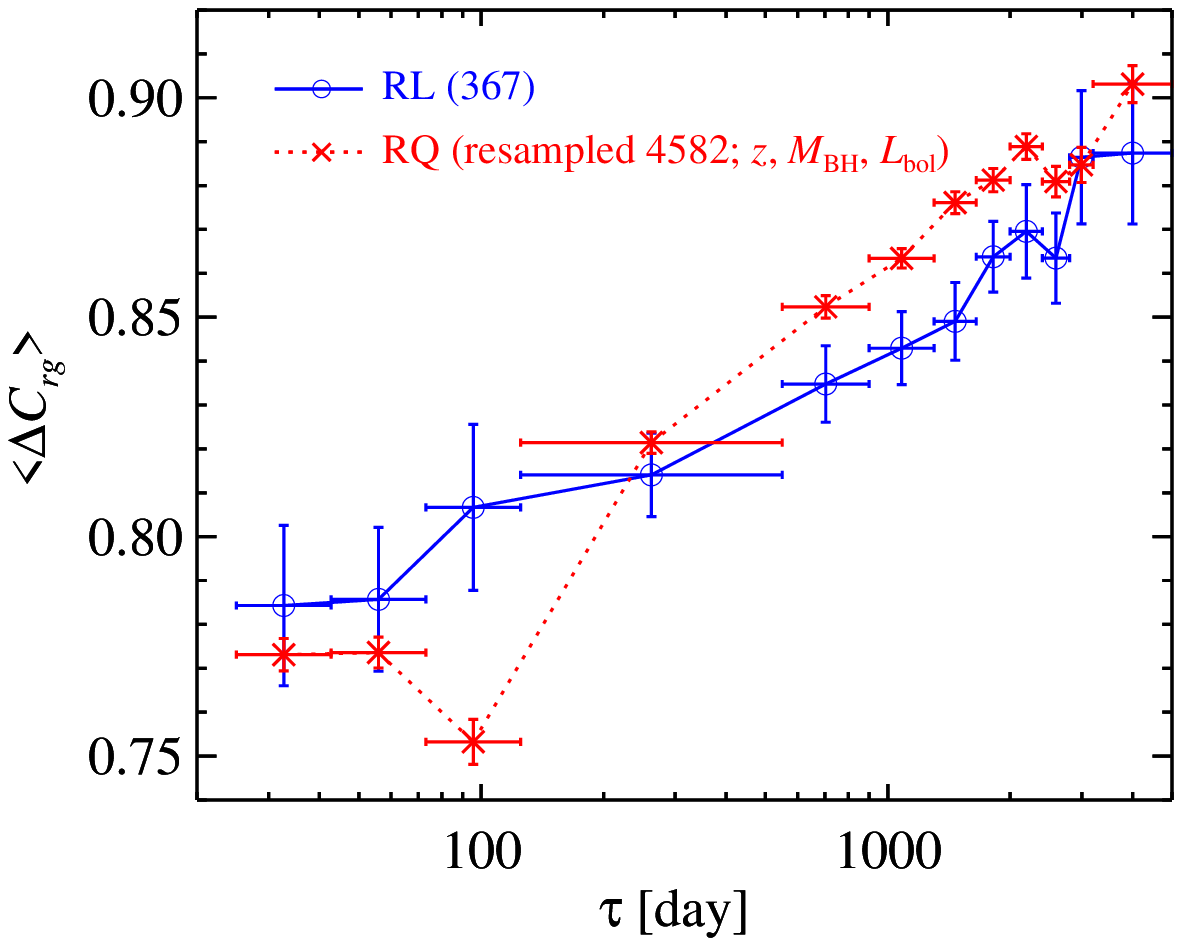}
	\includegraphics[width=0.39\textwidth]{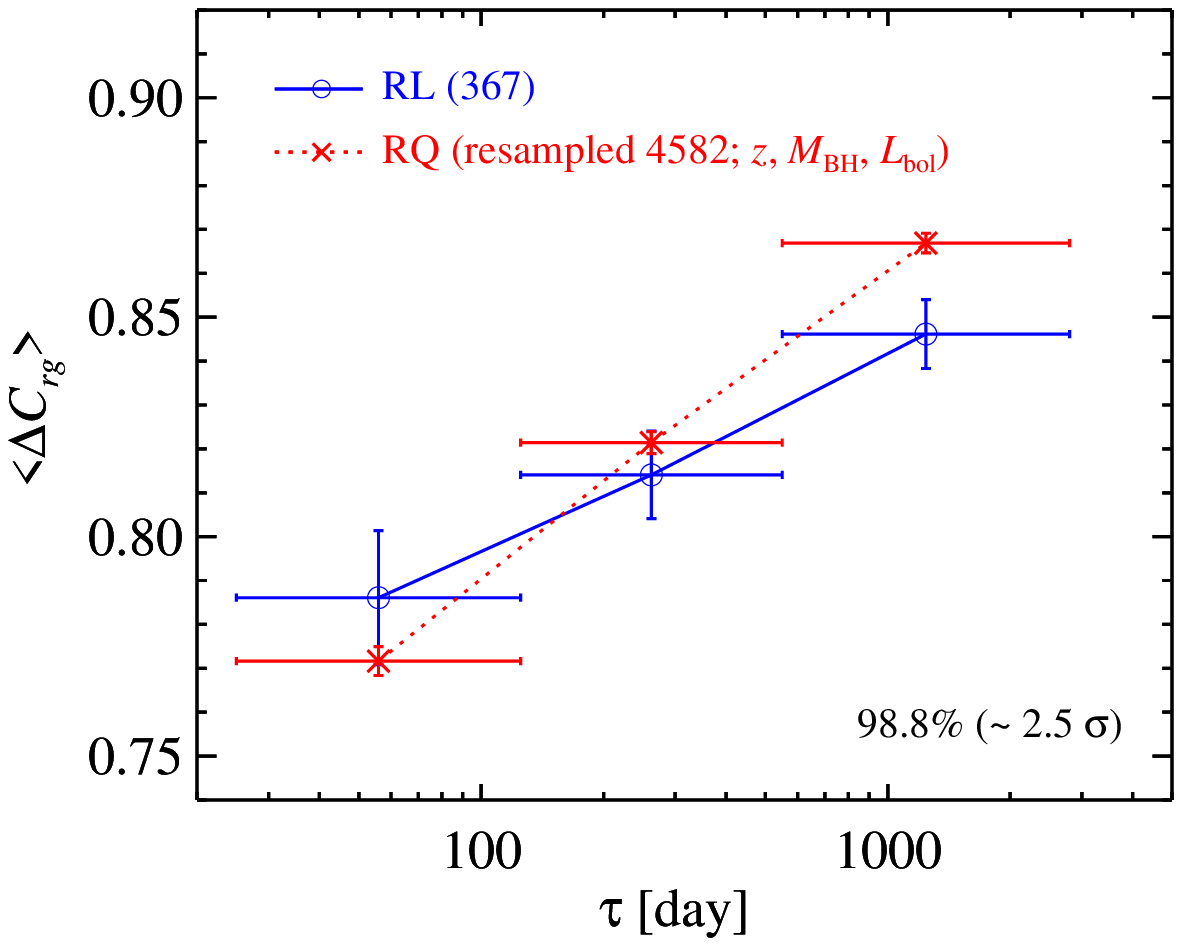}
	\caption{Similar as \cref{fig:cv_whole_sample}, but for the RL and resampled RQ sub-samples matched in $z$-$M_{\rm BH}$-$L_{\rm bol}$ (see Section~\ref{sect:resampled_z_m_Lb} and \cref{fig:dis_resampled_z_m_Lb}). }
	\label{fig:cv_resampled_z_m_Lb}
\end{figure}

\begin{figure}[H]
	\centering
	\includegraphics[width=0.39\textwidth]{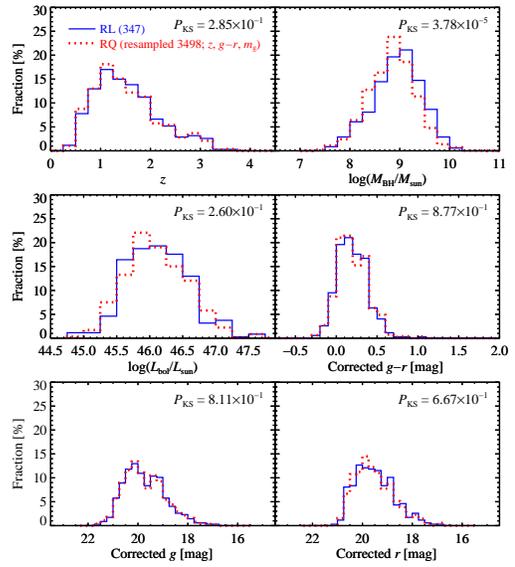}
	\caption{Same as \cref{fig:dis_whole_sample}, but for the RL and resampled RQ sub-samples matched in $z$, $g-r$ color, and $g$ magnitude. 
	(see Section~\ref{sect:resampled_z_gr_mg})
	}
	\label{fig:dis_resampled_z_gr_mg}
\end{figure}

As mentioned above, the RL sub-sample has redder $g-r$ color compared with the RQ sub-sample matched in $z$-$M_{\rm BH}$-$L_{\rm bol}$ (\cref{fig:dis_resampled_z_m_Lb}).
This is partly because RQ SDSS quasars were generally color selected, while RL ones were selected based on radio detections.
Redder colors could also lead to larger (smaller) photometric uncertainties in the $g$-band ($r$-band), and then may lead to a biased measurement of $\langle \Delta C_{rg} \rangle$.
To check whether our initial two populations are biased by these effects, we construct sub-samples for both RL and RQ sources matched in $z$, $g-r$ color, and $g$ magnitude, containing 347 and 3498 sources, whose relevant distributions are shown in \cref{fig:dis_resampled_z_gr_mg}.

Furthermore, since most of the quasars in Stripe 82 are selected according to the color-magnitude criteria proposed by 

\begin{figure}[H]
	\centering
	\includegraphics[width=0.39\textwidth]{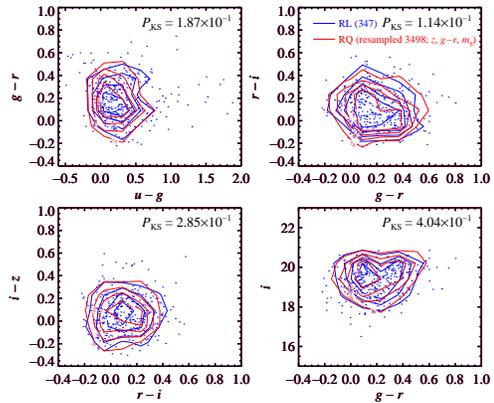}
	\caption{The color-color and magnitude-color diagrams for the RL (blue contour and dots for individual) and resampled RQ (red contour; missing individual for clarity) sub-samples matched in $z$, $g-r$ color, and $g$ magnitude. The two-dimensional K-S test suggests a similarity between these two sub-samples. (see Section~\ref{sect:resampled_z_gr_mg})
	}
	\label{fig:quasar_selection}
\end{figure}

\noindent Richards et al. \citet{Richards2002a} (see also Schmidt et al. \citep{Schmidt2010}), 
\cref{fig:quasar_selection} illustrates the same color-color and magnitude-color diagrams for our RL and resampled RQ sub-samples. 
Based on the 

\begin{figure}[H]
	\centering
	\includegraphics[width=0.39\textwidth]{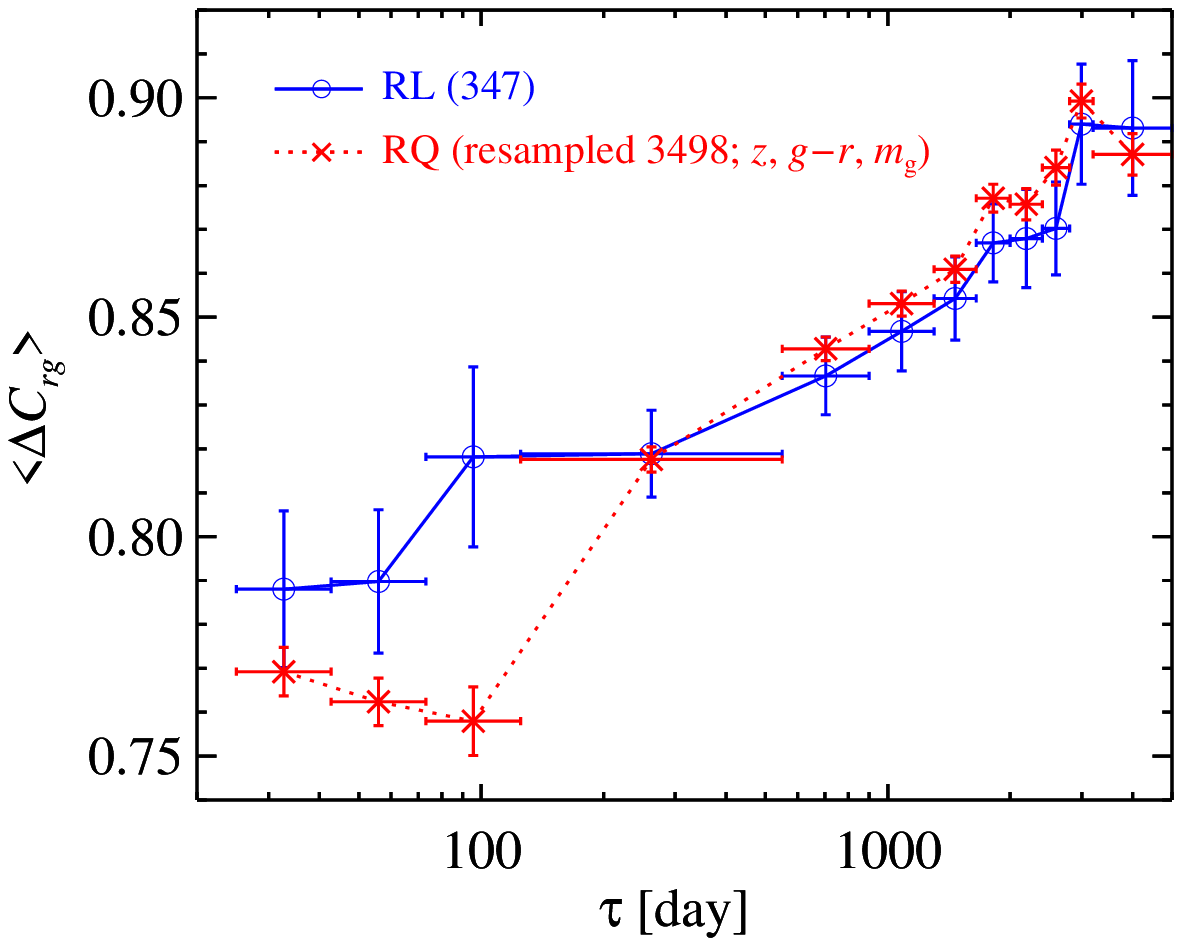}
	\includegraphics[width=0.39\textwidth]{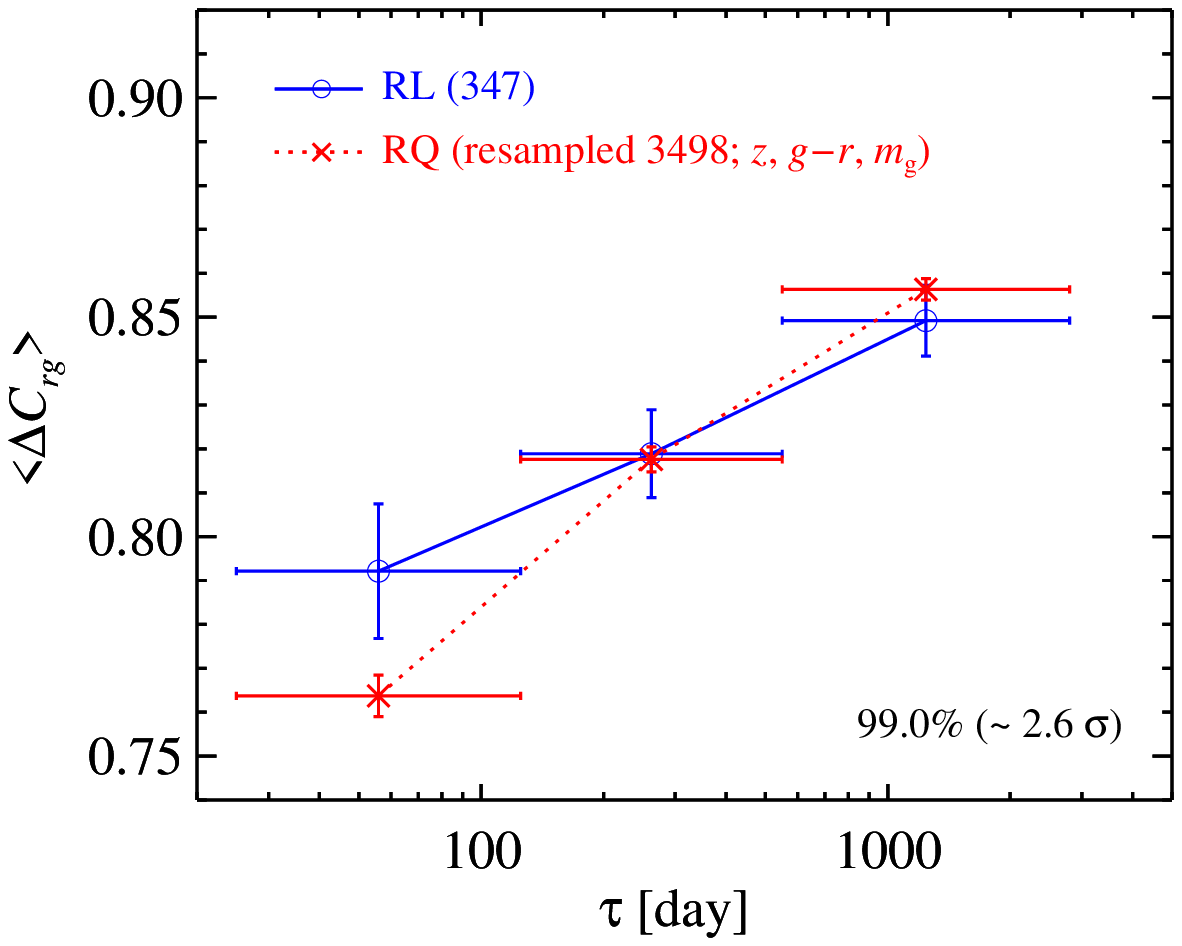}
	\caption{Similar as \cref{fig:cv_whole_sample}, but for the RL and resampled RQ sub-samples matched in $z$, $g-r$ color, and $g$ magnitude. (see Section~\ref{sect:resampled_z_gr_mg} and \cref{fig:dis_resampled_z_gr_mg}) }
	\label{fig:cv_resampled_z_gr_mg}
\end{figure}

\noindent two-dimensional K-S test, the similarity between these two-dimensional distributions of both sub-samples suggests that most of them would be selected in the same way using the same color criteria, and then the color selection bias could be negligible for these two sub-samples. These two sub-samples are then used to estimate the corresponding color variations, which are shown in \cref{fig:cv_resampled_z_gr_mg}.
The weaker/flatter timescale dependency for the RL sub-sample remains at a confidence level of $\sim 99.0\%$.
Therefore, our initial whole samples are not significantly biased by the color selection criteria, and then we do not restrict our initial sample using color selection not only because the complexity of quasar selection in 

\noindent Stripe 82 but also to keep as many sources as possible. Note that after matching in $z$, $g-r$ color, and $g$ magnitude, the initial source numbers reduce by $\simeq 16\%$ and 51\% for the RL and RQ samples, respectively.

\begin{figure}[H]
	\centering
	\includegraphics[width=0.39\textwidth]{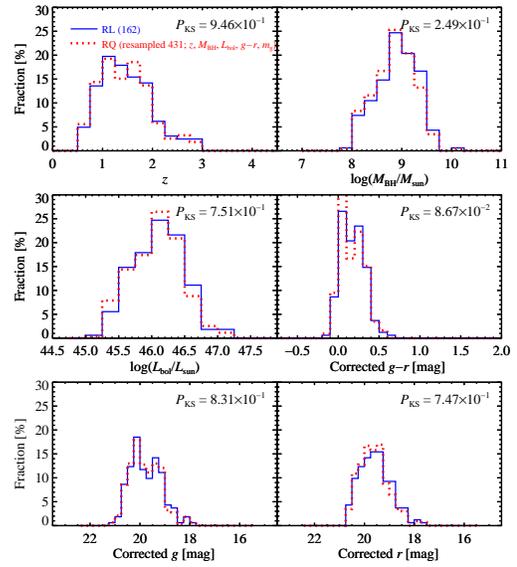}
	\caption{Same as \cref{fig:dis_whole_sample}, but for the RL and resampled RQ sub-samples matched in $z$, $M_{\rm BH}$, $L_{\rm bol}$, $g-r$ color, and $g$ magnitude. (see Section~\ref{sect:resampled_z_gr_mg})
	}
	\label{fig:dis_resampled_z_m_Lb_gr_mg}
\end{figure}

\begin{figure}[H]
	\centering
	\includegraphics[width=0.39\textwidth]{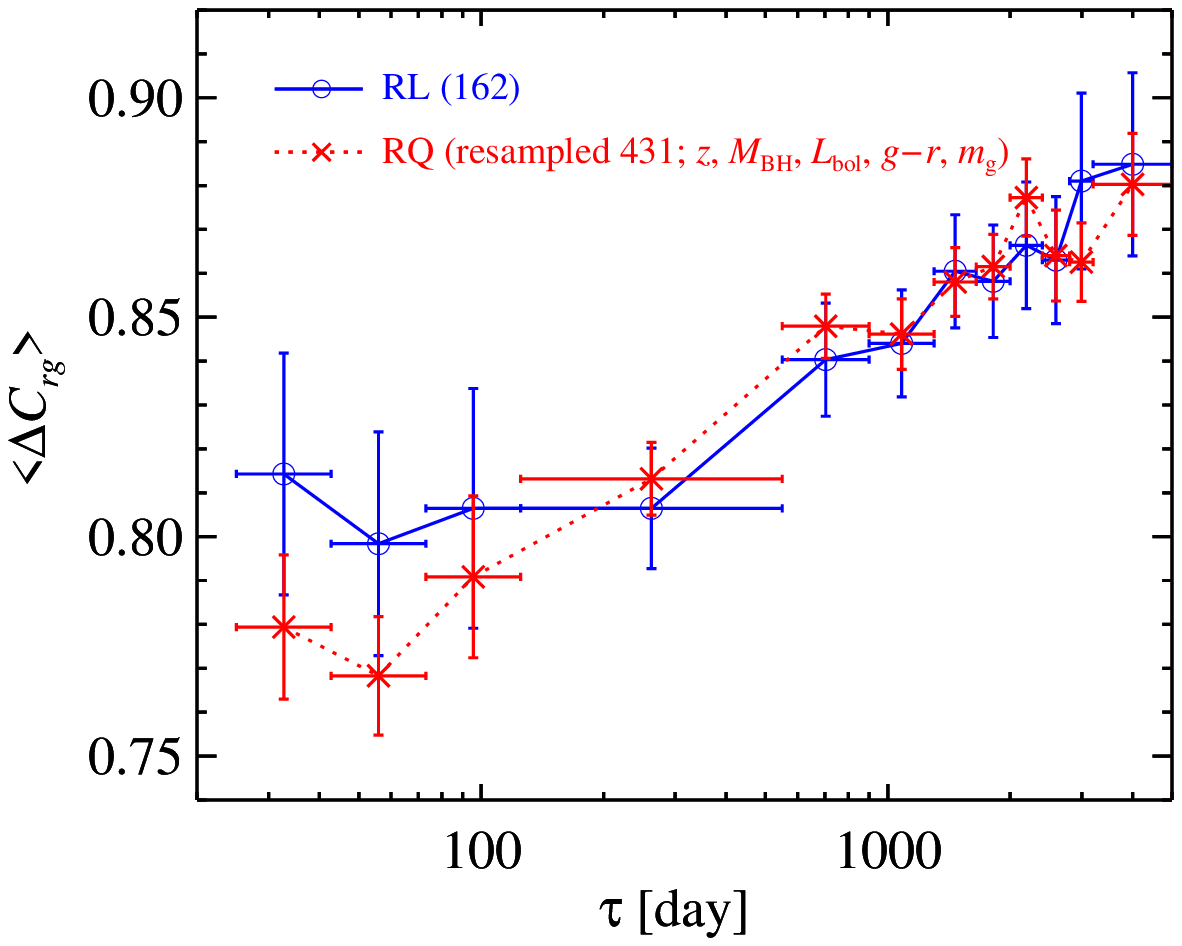}
	\includegraphics[width=0.39\textwidth]{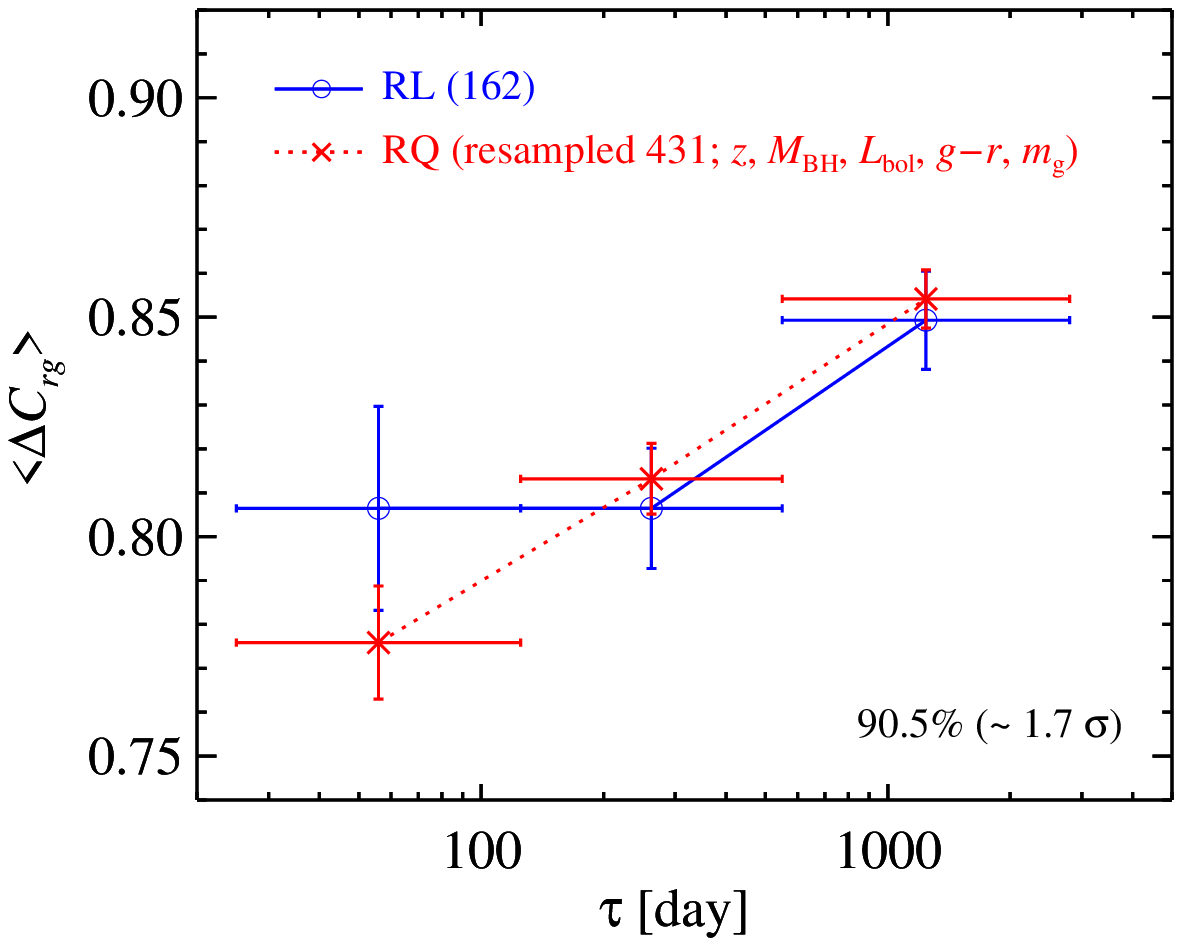}
	\caption{Similar as \cref{fig:cv_whole_sample}, but for the RL and resampled RQ sub-samples matched in $z$, $M_{\rm BH}$, $L_{\rm bol}$, $g-r$ color, and $g$ magnitude. (see Section~\ref{sect:resampled_z_gr_mg} and \cref{fig:dis_resampled_z_m_Lb_gr_mg}) }
	\label{fig:cv_resampled_z_m_Lb_gr_mg}
\end{figure}

One may have noticed the difference of the BH mass distribution between these two sub-samples from \cref{fig:dis_resampled_z_gr_mg}.
 Sub-samples for RL and RQ sources matched in $z$, $M_{\rm BH}$, $L_{\rm bol}$, $g-r$ color, and $g$ magnitude can be further constructed with the relevant distributions shown in \cref{fig:dis_resampled_z_m_Lb_gr_mg}.
Since the source numbers are significantly reduced, i.e., only 162 and 431 for the RL and RQ sub-samples, respectively, we can clearly expect that the resultant color variations possess larger uncertainties and the difference of color variation between RL and RQ sub-samples decreases to a lower confidence level as shown in \cref{fig:cv_resampled_z_m_Lb_gr_mg}.

\subsection{Excluding sources with BH mass inferred from \civ~emission line}\label{sect:no_CIV}

The BH masses given in the catalog of Shen et al. \citet{Shen2011} are viral products of the broad line width and the corresponding continuum luminosity
(i.e., H$\beta$ and $L_{5100 \angstrom}$ at $z< 0.7$, \mgii

\begin{figure}[H]
	\centering
	\includegraphics[width=0.39\textwidth]{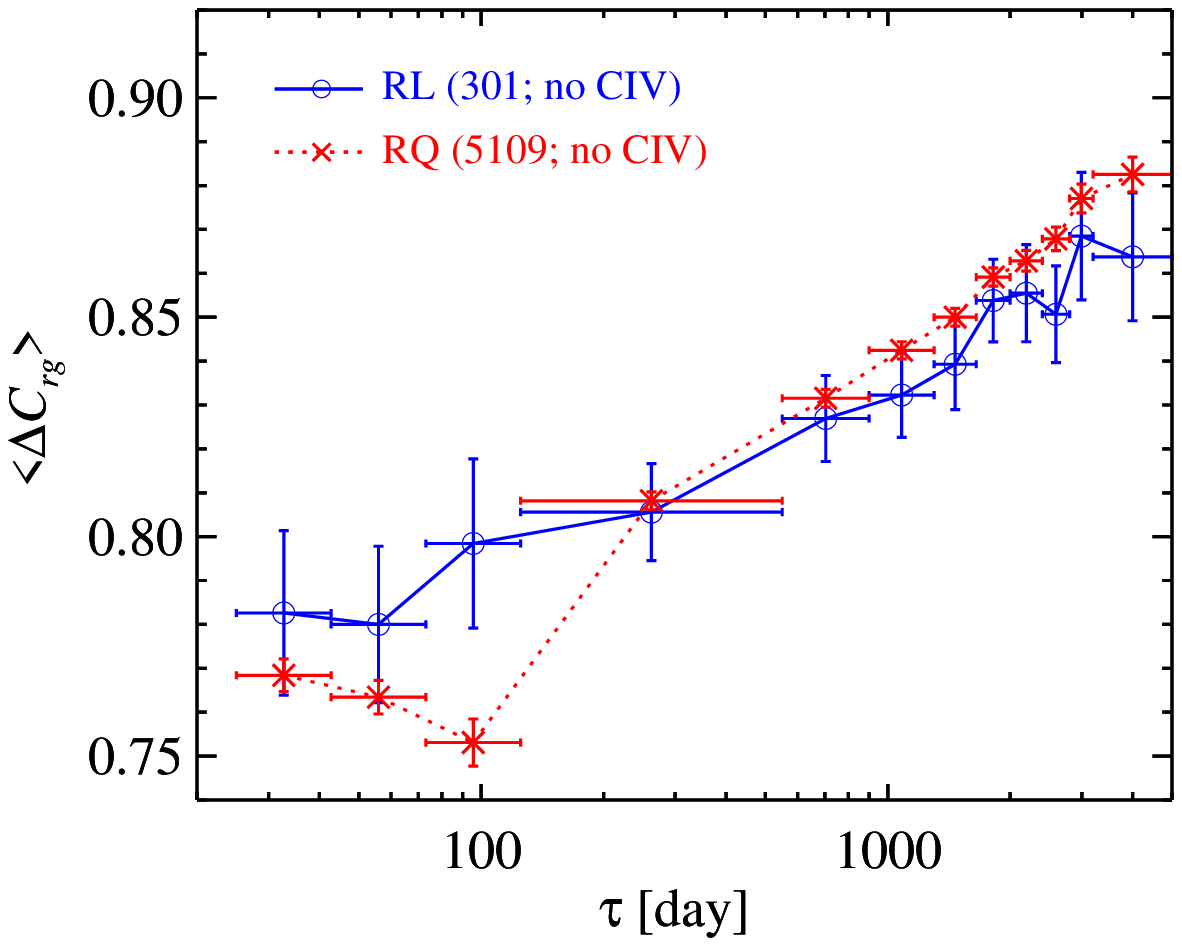}
	\includegraphics[width=0.39\textwidth]{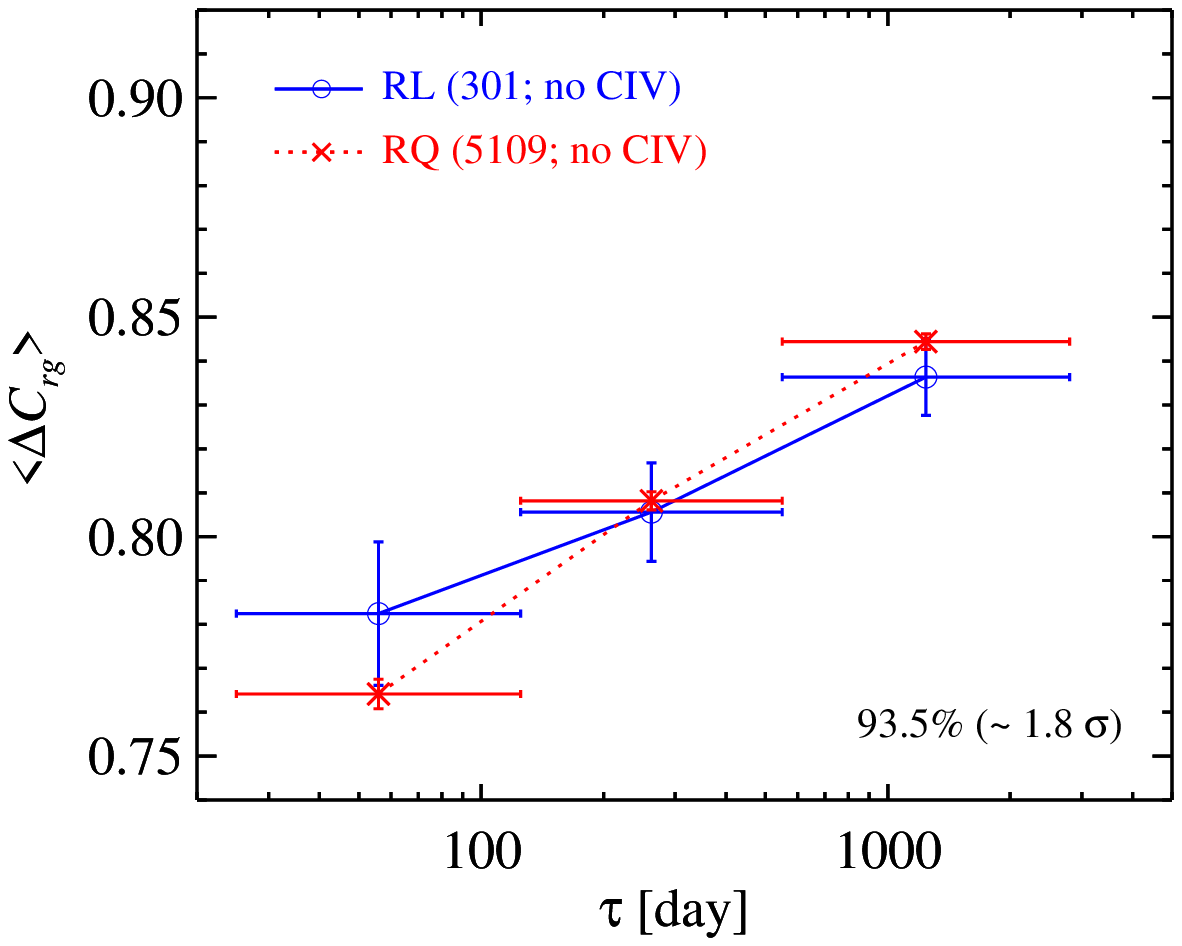}
	\caption{Same as \cref{fig:cv_whole_sample}, but for the RL and RQ sub-samples excluding sources with BH mass derived using \civ~emission line (primarily at $z \geq 1.9$). (see Section~\ref{sect:no_CIV}) }
	\label{fig:cv_noCIV}
\end{figure}

\noindent and $L_{3000 \angstrom}$ at $0.7 \leq z < 1.9$, and \civ~plus $L_{1350 \angstrom}$ at $z \geq 1.9$).
The bolometric luminosities were computed adopting bolometric corrections from the composite SED in Richards et al. \citet{Richards2006}.
We note that such bolometric luminosities and BH masses could be systematically biased for RL quasars because
1) for some RL quasars, the jet contribution to the observed continuum luminosities could be significant;
2) there is significant difference in the UV SED between RL and RQ quasars \citept[e.g.,][]{Francis2000,Ivezic2002,Kotilainen2007,Labita2008,Shankar2016}.
Furthermore, Coatman et al. \citet{Coatman2016,Coatman2017} have claimed that the BH mass derived using the velocity width of \civ~broad emission line, which exhibit significant blue asymmetry, can be overestimated by a factor of more than 5. Therefore, we construct sub-samples by excluding those sources with BH mass derived using \civ~emission line (primarily at $z \geq 1.9$).
The corresponding color variations of such sub-samples  are shown in \cref{fig:cv_noCIV}, where the confidence level at which the RL sub-sample shows a weaker/flatter timescale dependency slightly decreases to 93.5\%, likely due to the smaller sample sizes (the revised RL and RQ sub-samples contain 301 and 5109 sources, respectively, $\sim 70\%$ of the whole samples).
Much larger samples from future surveys would be useful to further confirm the difference in color variation between RL and RQ quasars.

\subsection{Structure Function}\label{sect:sf}

The structure function (SF) measures the variation amplitude as a function of time lag, $\tau$, between epochs.
Following MacLeod et al. \citet{MacLeod2012} (see e.g., Simonetti et al. \citep{Simonetti1985} and di Clemente et al. \citep{diClemente1996} for other forms of SF) to calculate the ensemble SFs of our samples in the $g$- and $r$-bands, we adopt
\begin{equation}
\mbox{SF}_{j} (\tau) = 0.74 \times {\rm IQR},
\end{equation}
where IQR is the 25\%-75\% interquartile range of the $j$-band $\Delta m^{\rm int}_j$ distribution. The difference of the intrinsic magnitude, $\Delta m^{\rm int}_j$, between two epochs (e.g., $t_1$ and $t_2$) is obtained by subtracting the corresponding photometric errors (e.g., $e_1$ and $e_2$) in quadrature from that of the observed magnitude, $\Delta m^{\rm obs}_j$, i.e., $\Delta m^{\rm int}_j = \sqrt{(\Delta m_j^{\rm obs})^2  - (e_1^2 + e_2^2) }$, where the radicand is set to zero if it is negative and kept to include the information of tiny variations.
This form is insensitive to outliers in the light curve due to poor photometric quality and is especially effective at short time lags where the intrinsic variations are weak \citep{MacLeod2012}. 

It is known that caution is required when inferring the potential characteristic timescales from the SF \citept[e.g.,][]{Emmanoulopoulos2010}. We stress that in this work we do not use the SFs to parameterize the quasar variations. Instead, we simply compare the ensemble SFs
of RL and RQ samples (i.e., to compare the typical 
variation amplitudes at different timescales). We further note that the light curves of these sources have similar sampling pattern (cf. \cref{fig:example}), validating the direct comparison.

In the top panel of \cref{fig:figsf}, we plot the ensemble SDSS $g$- and $r$-band SFs for the whole RL and RQ samples in the observer's frame. Greater amplitudes are found in the $g$-band (than in the $r$-band) for both RL and RQ quasars, consistent with former results that AGNs generally show larger variation amplitudes in bluer bands, or the BWB trend. 
The amplitude of $g$-band SF is comparable to that derived by MacLeod et al. \citet{MacLeod2012} (see their Figure~5).
We note that at long timescales ($>$ 500 days), RL and RQ quasars show similar variation amplitudes, consistent with previous studies \citept[e.g.,][]{Bauer2009,MacLeod2010,Zuo2012} which compared long term variation amplitudes of these two populations.
Interestingly, for the first time we see weaker variations in both the $g$- and $r$-bands (more prominent in the $g$-band) at shorter timescales ($\sim 25$ -- 300 days) in RL quasars, compared with those of RQ ones.

Following Zhu et al. \citet{Zhu2016}, we also obtain a $ \Delta C_{rg}^{\rm SF-ratio}$ -- $\tau$ relation directly inferred from the ratio of $r$- to $g$-band SFs, i.e., $\Delta C_{rg}^{\rm SF-ratio} \equiv {\rm SF}_r/{\rm SF}_g$, as illustrated in the lower panel of \cref{fig:figsf}. This plot is more or less consistent with \cref{fig:cv_whole_sample}, showing a potentially weaker/flatter relation for RL sources.
Note that to derive the color variation as shown in \cref{fig:cv_whole_sample} we have only selected those epoch pairs with statistically significant variations (see Section~\ref{sect:cv}), while the $\Delta C_{rg}^{\rm SF-ratio}$ is obtained from the ratio of SFs derived using all epoch pairs including a large number of those with variations dominated by photometric uncertainties. This could be the reason that the difference of the $ \Delta C_{rg}^{\rm SF-ratio}$ -- $\tau$ relation between RL and RQ samples shown in the bottom panel of \cref{fig:figsf} is less significant.

It is known that blazars show strong intra-night optical variability \citept[INOV; e.g.,][]{GuptaJoshi2005}, which seems in contrast to our finding of weaker variations in RL quasars at timescales of $\sim 25$ -- 300 days.
However, studies excluding blazers and highly optical polarized core-dominated quasars have shown similar duty cycles and INOV amplitudes between RL and RQ quasars \citep{Stalin2004a,Stalin2004b,Ramirez2009}.
Therefore, the weaker variations in our RL quasars at timescales of $\sim 25$ -- 300 days we detect does not necessarily contradict with INOV studies. Unfortunately, limited by photometric uncertainties, we are unable to extend the SFs to even shorter timescales.

\section{Discussion}\label{sect:discussion}

We have shown that both RL and RQ quasars show timescale dependence in their color variations, in a way that the shorter timescale variations are bluer than the longer term ones.
However, RL sources on average exhibit a potentially weaker/flatter dependence on timescale.

Based on the inhomogeneous accretion disk model firstly developed by Dexter \& Agol \citet{DexterAgol2011}, Cai et al. \citet{Cai2016} have demonstrated that a revised thermal-fluctuating disk model can well recover the observed timescale dependence of the color variation observed in quasars \citep{Sun2014}.
While the observed timescale dependences in both populations suggest the variations could be due to thermal fluctuations in the accretion disk
\citep{Cai2016,Cai2018}, the different dependence of color variation between RL and RQ populations may hint that the inner disk properties are distinct, and can be investigated through inhomogeneous disk simulations.

\subsection{Simulations with inhomogeneous disk models}\label{sect:model}

A standard thin disk model (surrounding a non-rotating BH) is assumed by Cai et al. \citet{Cai2016} to simulate the thermal fluctuations in accretion disk.
The disk is split into individual zones, the temperatures of which (in logarithm space) are subject to random and independent fluctuations following damped random walk process \citep{Kelly2009}.
Such damped random walk process is described by two key parameters, which are the characteristic timescale, $\tau$, and the variation amplitude at infinite timescale,

\begin{figure}[H]
	\centering
	\includegraphics[width=0.39\textwidth]{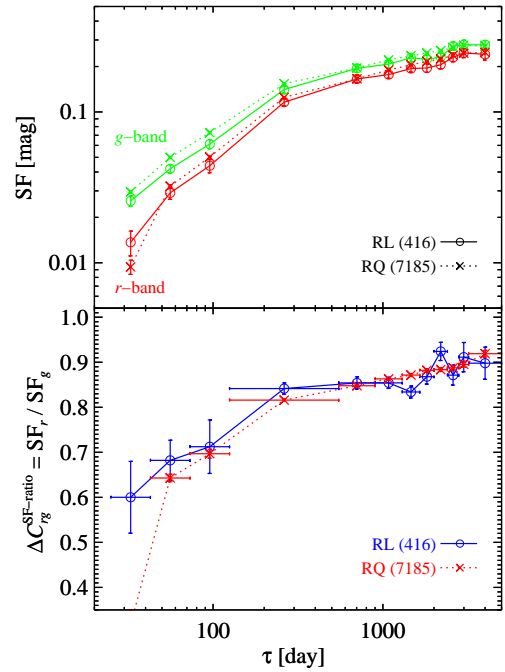}
	\caption{Top panel: The $g$- (green) and $r$-band (red) SFs for the whole RL (solid line) and RQ (dotted line) samples. Bottom panel: The color variations for the whole RL (blue solid line) and RQ (red dotted line) samples directly inferred from the ratio of $r$- to $g$-band SFs as shown in the upper panel. (see Section~\ref{sect:sf})
	}
	\label{fig:figsf}
\end{figure}

\noindent $\sigma_{\rm l}$.
As assumed by Cai et al. \citet{Cai2016}, the characteristic timescales of the thermal fluctuations in the disk is radius-dependent (i.e., $\tau \sim r$), while $\sigma_{\rm l}$ is constant.

Below we consider four different diagrams to investigate what kind of revisions to the inhomogeneous thin disk model of Cai et al. \citet{Cai2016} can yield a weaker/flatter timescale dependence as we observed in RL quasars:

A) The accretion disk in RL quasar is truncated at a break radius, $r_{\rm br}$;

B) The inner accretion disk in RL quasar is cooler compared with that in RQ one, following
\begin{equation}
  \log T^{\rm RL}(r) = \log T^{\rm RQ}(r) - A_{\rm T} \times \exp( -{r}/{r_{*}} );
\end{equation}

C) The inner accretion disk in RL source is more stable with smaller $\sigma_{\rm l}$:
\begin{equation}
\sigma_{\rm l}^{\rm RL}(r) / \sigma_{\rm l}^{\rm RQ}(r) = 1 - A_{\sigma} \times \exp( -{r}/{r_{*}} );
\end{equation}

D) The inner accretion disk in RL quasar fluctuates more slowly with larger $\tau$:
\begin{equation}
\tau^{\rm RL}(r) / \tau^{\rm RQ}(r) = 1 + A_{\tau} \times \exp( -{r}/{r_{*}} );
\end{equation}
where $r_{*}$ is the typical transition radius in each model, and $A_i$ the adjusted amplitude corresponding to its subscript, $i$.

\begin{figure*}
	\centering
	\includegraphics[scale=0.5]{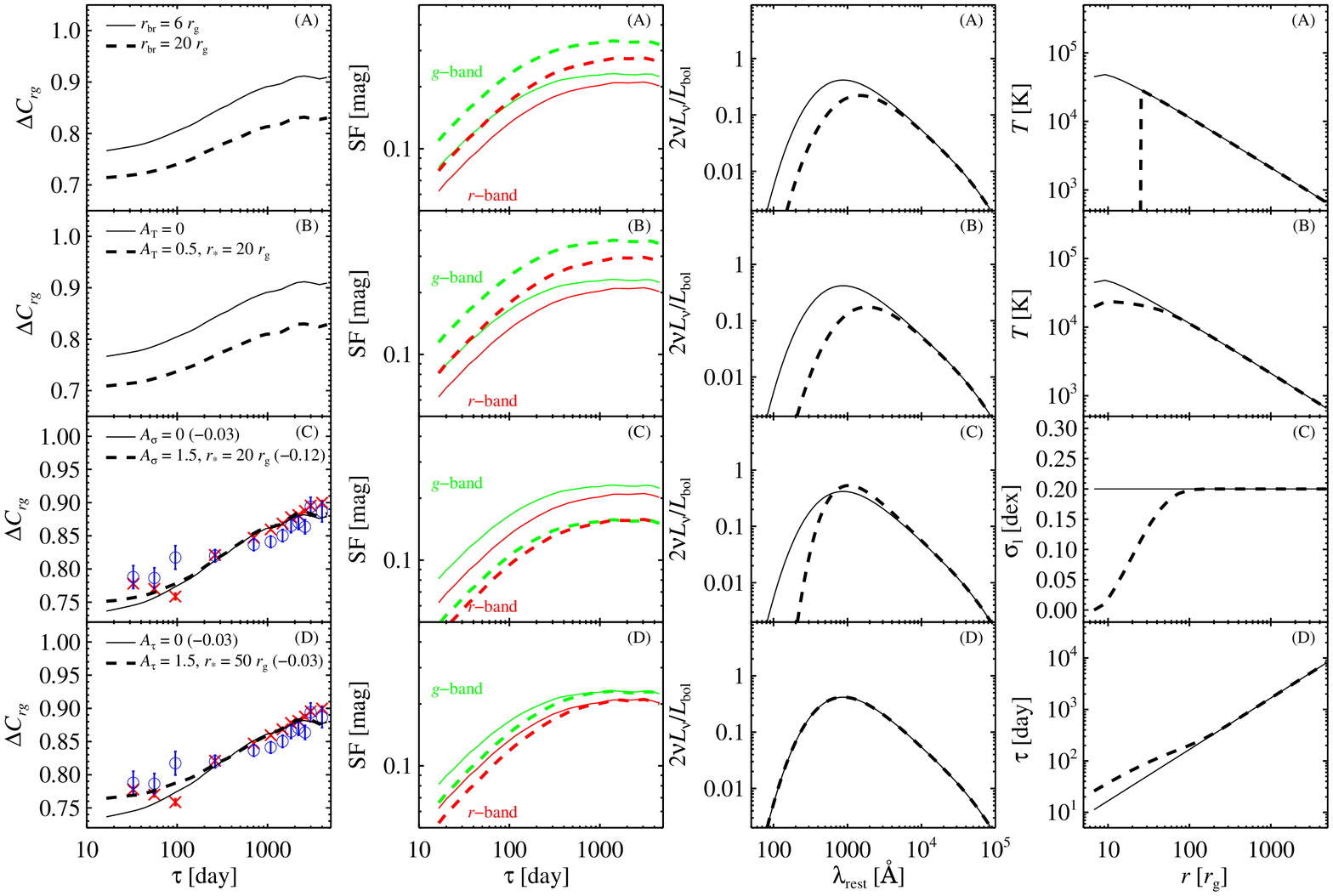}
	\caption{The simulated $\Delta C_{rg}$ -- $\tau$ relations (left column), SFs (middle-left column), mean SEDs (middle-right column), and the relevant properties as a function of radius (right column), compared between the reference inhomogeneous accretion disk model of Cai et al. \citep{Cai2016} ($r_{\rm br} = 6~r_{\rm g}$, $A_{\rm T} = 0$, $A_\sigma =0$, and $A_\tau = 0$) for RQ quasars (thin solid lines) and our four adjusted models for RL quasars (thick dashed lines): 
	(A) The accretion disk truncated at a break radius, $r_{\rm br} = 20~r_{\rm g}$; (B) Cooler disk temperature at inner disk ($A_{\rm T} = 0.5$ with $r_* = 20~r_{\rm g}$); (C) Smaller variation amplitude, $\sigma_{\rm l}$, at inner disk ($A_\sigma = 1.5$ with $r_* = 20~r_{\rm g}$); (D) Slower temperature fluctuation at inner disk ($A_\tau = 1.5$ with $r_* = 50~r_{\rm g}$). For comparison, also shown are the observed $\langle \Delta C_{rg} \rangle$ -- $\tau$ relations from \cref{fig:cv_whole_sample} (blue open circles and red crosses for the whole RL and RQ samples, respectively), and then the simulated relations are also vertically shifted downward with negative numbers nominated within the corresponding parentheses (see Section~\ref{sect:model}).
	}
	\label{figsim}
\end{figure*}

In each paradigm, other model settings are kept identical for RL and RQ quasars. Following Cai et al. \citet{Cai2016}, we simulate the $g$- and $r$-band light curves for each of these diagrams,
adopting redshift $z = 1.7$, BH mass $M_{\rm BH} = 5\times10^{8}~M_{\rm sun}$, and accretion rate $\dot{M} = 1~M_{\rm sun}~{\rm yr}^{-1}$. The characteristic timescale at the innermost radius, i.e., $6~r_{\rm g}$, of a thin disk is set to 10 days and
the variation amplitude $\sigma_{\rm l}$ to 0.2 dex (if not revised for RL quasars accordingly).
For each set of parameters, we simulate the $g$- and $r$-band light curves in step of 1 day in the rest-frame for 100 sources with an observed time length of 10 years, similar to that of SDSS quasars. Furthermore, we have used the same random numbers among different sets of parameters in order to highlight the effects of the corresponding parameters.
Averaging over these simulations, we draw in \cref{figsim} the simulated $\Delta C_{rg}$ -- $\tau$ relation, the $g$-/$r$-band SFs, the mean disk SEDs, and the relevant properties as a function of the disk radius.
In the lower left two panels, the observed data from \cref{fig:cv_whole_sample} are over-plotted for comparison; the models are also shifted vertically following Cai et al. \citep{Cai2016} to match the observed amplitudes of $\Delta C_{rg}$ at longer timescales.

We see that, while model A and model B can reproduce redder mean SEDs, they both predict stronger variations in amplitude\footnote{The stronger variation is due to the fact that in model A and model B, there are less disk zones which contribute to observed emission in the $g$- and $r$-bands. More disk zones fluctuating independently would naturally reduce the variation amplitude of the integrated emission \citep{Cai2016}.}
(larger SFs) and much bluer color variations (smaller $\Delta C_{rg}$) with identical tendencies,
which appear inconsistent with our observations. 
A marginally flatter relation, a globally smaller SF, and a bluer mean SED are implied by model C, which are partly consistent with our observations.
Interestingly, model D can yield a flatter $\Delta C_{rg}$ -- $\tau$ relation, and smaller variation amplitudes (only at shorter timescales), well consistent with our discoveries. 
Meanwhile, model D does not alter the mean SED\footnote{Stronger fluctuations would yield bluer mean SEDs as each disk zone is more likely to have higher temperature, while the total emission from the disk averaged over a long time interval remains unchanged \citep{Cai2016}.}.
Both model C and model D correspond to more stable inner accretion either with smaller variation amplitude or slower variation.
We note that while model A or model B alone is inconsistent with our observation, this study does not necessarily rule out the paradigms in these models as they might be at work together with model C and model D.

Therefore, such comparison suggests that the inner accretion disks in RL quasars may be more stable, with smaller fluctuation amplitudes and possibly with also longer characteristic timescales,
compared with those in the RQ ones with similar redshifts, BH masses, and bolometric luminosities.
Note that, besides these differences in the inner disk, the difference of mean SEDs between RL and RQ quasars may be also attributed to their intrinsic difference of BH spins which introduces different innermost stable circular orbits. The difference in BH spin between RL and RQ quasars has also been suggested by various studies \citept[e.g.,][]{KratzerRichards2015,Schulze2017b}. Further detailed comparisons and discussions will be delayed for future work.

\subsection{Alternative possibilities?}

In additional to disk fluctuations, there are other factors which may contribute to the observed UV/optical variations in quasars.
For RL sources, the jet contamination might be non-negligible.
A possible logic to link the jet contamination to the flatter timescale dependence of the color variation in RL quasars is that the jet component, redder than the disk emission, may contribute more to the observed variations at shorter timescales, thus yielding a flatter $\Delta C_{rg}$ -- $\tau$ relation. However, we do not see stronger variations in our RL sources at shorter timescales compared with the RQ ones, and the observation indeed shows
an opposite trend  (see \cref{fig:figsf}).

Furthermore, according to MacLeod et al. \citet{MacLeod2010}, only quasars with highest radio loudness ($R > 1000$) show stronger variations compared with the RQ ones, suggesting that the jet contamination to the UV/optical emission and variation is minor for most RL quasars.
Excluding quasars with radio loudness $R > 1000$ (79 of 416 RL sources) from our analyses does not alter the results we present in this work.

The X-ray reprocessing may also play a role in producing the UV/optical variations in quasars.
It is found that RL quasars tend to have higher X-ray-to-optical ratios \citept[e.g.,][]{Worrall1987,Ballo2012}, thus in this population the X-ray reprocessing could be more important and
may lead to a flatter $\Delta C_{rg}$ -- $\tau$ relation as reprocessing could yield timescale-independent color variation \citep{Zhu2018}.
However, this possibility is unlikely either as we do not see stronger variations in RL quasars, particularly at shorter timescales.
Besides, as found by Miller et al. \citet{Miller2011}, the radio-intermediate quasars are only modestly X-ray bright relative to RQ quasars, and only RL quasars with high radio luminosities and $R \gtrsim 3000$ (28 out of 416 in the whole RL sample) become strongly X-ray bright.
Furthermore, stronger X-ray emission in RL sources may not necessarily lead to stronger X-ray reprocessing, considering the X-ray corona
could be relativistically outflowing \citept[e.g.,][]{Liu2014,King2017,ChainakunYoung2017}.

\subsection{Clues to Jet Launching in quasars}

Our inhomogeneous disk simulations suggest that the inner accretion disks in RL quasars may be more stable, compared with those in RQ ones.
The physical origin of the UV/optical disk fluctuations is still poorly known. Interestingly, theoretical works show that the thin accretion disks could be stabilized by strong magnetic fields \citept[e.g.,][]{BegelmanPringle2007,Sadowski2016c,Sadowski2016d,Zheng2011}.
Meanwhile, a strong magnetic field is commonly believed to be essential for jet launching \citep{BlandfordPayne1982,BlandfordZnajek1977,Tchekhovskoy2011}. Such strong magnetic fields have been detected in the jet bases of some individual RL AGNs
\citept[e.g.,][]{Marti-Vidal2016, Baczko2016}, though we have little knowledge on the strength of magnetic field in RQ ones.
The differences between RL and RQ quasars we present in this work therefore suggest RL quasars have stronger magnetic field in their inner region, which makes the disk more stable, and is one of the key underlying differences between the two populations.

The suppression of disk fluctuations by the stronger magnetic fields in RL quasars  could be more prominent than what we observed in this work, considering jet could also contribute to the observed variations.

We remark that while our simulations of inhomogeneous disk emission are helpful to interpret the observations we present in this work, such toy models (with modifications to thin disk) are still rather simple.
Particularly, we simply assume non-rotating BHs, which could be improper as BH spin could be a key factor behind the jet launching in quasars \citept[e.g.,][]{BlandfordZnajek1977,Tchekhovskoy2010}. Thin disk surrounding fast rotating BH can extend to smaller inner radii than what we adopted in this work (i.e., 6 $r_{\rm g}$).  Therefore, the fact that our mode A and model B are incompatible to the observations does not necessarily rule out the possibility that the inner accretion disks in RL quasars are truncated or cooler (within much smaller radius) than those in RQ ones. Extensive simulations considering the effect of BH spin will be carried out in the near future.

Nevertheless, the findings of this investigation indicate that the inner accretion disks in RL quasars may be more stable than those in RQ ones. The difference could be due to the stronger magnetic field in the RL quasars, which is essential to both jet launching and inner disk stabilization.

\Acknowledgements{We acknowledge the anonymous referee for valuable comments. We thank Xinwu Cao for helpful discussion.
This work is supported by National Basic Research Program of China (973 program, grant No. 2015CB857005) and National Science Foundation of China (grants No. 11233002, 11421303, 11503024 $\&$ 11873045). 
J.X.W. thanks support from Chinese Top-notch Young Talents Program, and CAS Frontier Science Key Research Program (QYZDJ-SSW-SLH006).
Z.Y.C. acknowledges support from the Fundamental Research Funds for the Central Universities.
F.Y. thanks the grant from the Ministry of Science and Technology of China (No. 2016YFA0400704).}

\InterestConflict{The authors declare that they have no conflict of interest.}


\bibliography{ms.bbl}
\bibliographystyle{scpma}

\end{multicols}

\end{document}